\documentclass[journal]{IEEEtran}
\IEEEoverridecommandlockouts
\usepackage{cite}
\usepackage{amsmath,amssymb,amsfonts}
\usepackage{algorithmic}
\usepackage{graphicx}
\usepackage{subfig}
\usepackage{textcomp}
\usepackage{xcolor}
\usepackage{acronym}
\usepackage{cleveref}
\usepackage{pdflscape}

\acrodef{uc}[UC]{unit commitment}
\acrodef{ed}[ED]{economic dispatch}
\acrodef{so}[SO]{system operator}
\acrodef{ufls}[UFLS]{underfrequency load shedding}
\acrodef{ofgd}[OFGD]{overfrequency generator disconnection}
\acrodef{rocof}[ROCOF]{rate of change of frequency}
\acrodef{res}[RES]{renewable energy source}
\acrodef{milp}[MILP]{mixed integer linear program}
\acrodef{fcuc}[FCUC]{frequency-constrained \acs{uc}}
\acrodef{pfcuc}[P-FCUC]{preventive frequency-constrained \acs{uc}}
\acrodef{cfcuc}[C-FCUC]{corrective frequency-constrained \acs{uc}}
\acrodef{suc}[SUC]{security-constrained \acs{uc}}
\acrodef{ruc}[RUC]{Robust \acs{uc}}
\acrodef{sfr}[SFR]{system frequency response}
\acrodef{vll}[VLL]{value of lost load}

\usepackage[intoc]{nomencl}
\usepackage{xstring}
\usepackage{xpatch}
\usepackage{eurosym}

\acrodef{uc}[UC]{unit commitment}
\acrodef{ed}[ED]{economic dispatch}
\acrodef{so}[SO]{system operator}
\acrodef{ufls}[UFLS]{underfrequency load shedding}
\acrodef{ofgd}[OFGD]{overfrequency generator disconnection}
\acrodef{rocof}[ROCOF]{rate of change of frequency}
\acrodef{res}[RES]{renewable energy source}
\acrodef{milp}[MILP]{mixed integer linear program}
\acrodef{fcuc}[FCUC]{frequency-constrained \acs{uc}}
\acrodef{pfcuc}[P-FCUC]{preventive frequency-constrained \acs{uc}}
\acrodef{cfcuc}[C-FCUC]{corrective frequency-constrained \acs{uc}}
\acrodef{suc}[SUC]{security-constrained \acs{uc}}
\acrodef{ruc}[RUC]{Robust \acs{uc}}
\acrodef{sfr}[SFR]{system frequency response}
\acrodef{vll}[VLL]{value of lost load}

\patchcmd{\thenomenclature}
  {\leftmargin\labelwidth}
  {\leftmargin\labelwidth\itemindent 1em }
  {}{}
\newcommand{\nomenclheader}[1]{%
  \item[\hspace*{-\itemindent}\normalfont\bfseries#1]}
\renewcommand\nomgroup[1]{%
  \IfStrEqCase{#1}{%
   {G}{\nomenclheader{Standard \acs{uc}:}}
   {F}{\nomenclheader{\acs{cfcuc}:}}
  }%
}

\makenomenclature

\nomenclature[G]{\(suc(.)\)}{Start-up costs [\euro]}
\nomenclature[G]{\(gc(.)\)}{Generation costs [\euro]}
\nomenclature[G]{\(x\)}{Commitment variable [$\in$\{0,1\}]}
\nomenclature[G]{\(P\)}{Power variable [MW]}
\nomenclature[G]{\(t\)}{Index of time intervals}
\nomenclature[G]{\(\mathcal{T}\)}{Set of all time intervals}
\nomenclature[G]{\(g\)}{Index of generators}
\nomenclature[G]{\(\mathcal{G}\)}{Set of all generators}
\nomenclature[G]{\(tt\)}{Alias index for time intervals}
\nomenclature[G]{\(gg\)}{Alias index for generators}
\nomenclature[G]{\(\overline{\mathcal{P}}_g\)}{Maximum power output of generator $g$ [MW]}
\nomenclature[G]{\(D\)}{Demand [MW]}
\nomenclature[G]{\(L_g\)}{Loss factor [-]}
\nomenclature[F]{\(\Delta F_{nadir}\)}{Allowable nadir frequency deviation [pu]}
\nomenclature[F]{\(K_g\)}{Governor gain of generator $g$ [pu]}
\nomenclature[F]{\(T_g\)}{Turbine-governor system time constant of generator $g$  [s]}
\nomenclature[F]{\(S_{base}\)}{System rating [MVA]}
\nomenclature[F]{\(H_g\)}{Inertia of generator $g$ [s]}
\nomenclature[F]{\(lsc(.)\)}{\acs{ufls} cost [\euro]}
\nomenclature[F]{\(C_{ufls}\)}{Linear \acs{ufls} cost [\euro/MW]}

\def\BibTeX{{\rm B\kern-.05em{\sc i\kern-.025em b}\kern-.08em
    T\kern-.1667em\lower.7ex\hbox{E}\kern-.125emX}}
    
\begin{document}

\title{Unit commitment with analytical underfrequency load-shedding constraints for island power systems
\thanks{Grant RTI2018-100965-A-I00 funded by MCIN/AEI/ 10.13039/501100011033 and by “ERDF A way of making Europe”.}
}

\author{Almudena~Rouco, Mohammad~Rajabdorri, Lukas~Sigrist,~\IEEEmembership{Member,~IEEE
}, Enrique~Lobato, Ignacio~Egido}

\maketitle

\begin{abstract}
This letter presents a \ac{cfcuc} for island power systems implementing analytical constraints on \ac{ufls}. Since \ac{ufls} is inevitable for sufficiently large disturbances, one can argue that less spinning reserve could be held back since \ac{ufls} takes place anyway. Congruently, the reserve criterion should consider \ac{ufls} likely to occur under disturbances. The \ac{cfcuc} can be converted into a \ac{pfcuc} or the standard \ac{uc} and the \ac{cfcuc} is thus a generalization. The proposed formulation is successfully applied to a Spanish island power system.
\end{abstract}


\printnomenclature

\section{Introduction}
Operation planning of island power systems is majorly carried out in a centralized manner, i.e., the \ac{so} sequentially carries out \ac{uc} and \ac{ed} over different time horizons. Optimal operation planning is restricted by, among others, the security of supply constraints. Typically, the security of supply is guaranteed by requiring a certain amount of spinning reserve, which is activated when an unplanned disturbance occurs. 

Currently, the reserve criterion is static and such that the amount of available reserve is sufficient to cover expected unplanned disturbances in terms of power and energy needed \cite{rajabdorri2022robust}. The timely activation of the reserve is however not guaranteed. Whereas frequency deviations due to smaller disturbances can be only confined thanks to the primary frequency control, \ac{ufls} or \ac{ofgd} take place to arrest frequency deviations under very large and even moderate active power unbalances. Since \ac{ufls} is inevitable for sufficiently large disturbances, one can argue that less spinning reserve could be held back since \ac{ufls} takes place anyway. Congruently, the reserve criterion should consider \ac{ufls} likely to occur under disturbances. The objective is not to increase \ac{ufls} unnecessarily but to reduce the amount of required spinning reserve to be available by acknowledging that sufficiently severe disturbances lead to \ac{ufls}.

The literature has already paid attention to the timely activation of the reserve under the framework of \ac{pfcuc}. \ac{pfcuc} approaches can be broadly grouped into analytical or data-driven approaches \cite{rajabdorri2022robust}. The former includes explicit expressions for \ac{rocof}, nadir, and steady-state frequency \cite{ferrandon2022inclusion, badesa2019simultaneous, shahidehpour2021two}, whereas the latter estimates them through functions deduced from a large training set. Reference \cite{rajabdorri2022robust} contains an up to date state of the art of both analytical and data-driven \ac{pfcuc}. The inclusion of constraints describing potential \ac{ufls}, leading to a \ac{cfcuc} approach, has however not been reported to the knowledge of the authors. This letter fills in the gap by formulating analytical constraints of \ac{ufls}. The proposed \ac{cfcuc} approach becomes a \ac{pfcuc} approach for large \ac{ufls} costs, and the standard \ac{uc} by omitting the \ac{ufls}-related constraints. 

This letter is organized as follows: first, the standard \ac{uc} formulation is briefly reviewed in section \ref{sec:stuc}. The formulation of the proposed \ac{cfcuc} is presented in section \ref{sec:cfcuc}. \Ref{sec:results} applies the proposed \ac{cfcuc} to a small Spanish island power system. \Ref{sec:conc} concludes the paper.

\section{Standard \ac{uc} formulation}

\label{sec:stuc}
The operation planning problem is formulated as a \ac{uc}, minimizing variable operation costs for a given horizon. The \ac{uc} determines the hourly program of generating units for a given horizon (e.g., weekly \ac{uc}, daily \ac{uc}, etc.). In particular, generation set points and start-up and shut-downs are determined. The standard \ac{uc} is formulated as a \ac{milp} problem \cite{rajabdorri2022robust}.

\begin{subequations} \label{equ1}
\begin{align}
\min\limits_{x,P} suc(x_{g,t})&+gc(P_{g,t}) \tag{\ref{equ1}}
\\
\sum\limits_{g\in\mathcal{G}, g\neq gg}\Big(\overline{\mathcal{P}}_{g,t} \cdot x_{g,t}-P_{g,t}\Big)&\geq L_{gg} \cdot P_{gg,t}&&\text{\scriptsize $t\in\mathcal{T},\;gg\in\mathcal{G}$}\label{equ1k}
\end{align}
\end{subequations}

The aim is to solve \eqref{equ1} subject to a set of constraints. For simplicity and without loss of generality, only the security of supply constraints are shown, whereas constraints on demand balance, commitment status, power output, and power ramping are omitted. $gc(.)$ is usually a quadratic cost function, which will be piecewise linearized. \Cref{equ1k} is the current reserve constraint and makes sure that there is enough reserve to compensate for the active power disturbance of generating unit $gg$. A priori, $L_{gg}$ takes the value of 1.0. In the case of modeling a single equivalent \acs{res} generation, $L_{gg}$ represents the expected fraction of \acs{res} generation to be lost.

\section{Formulation of the \ac{cfcuc}}

\label{sec:cfcuc}

The proposed \ac{cfcuc} estimates the necessary \ac{ufls} after a disturbance. Short-term frequency dynamics mostly depend on the inertia (physical or emulated) and turbine-governor systems. The critical size of the outage of generating unit $gg$, $P_{crit,gg,t}$, that causes the nadir frequency deviation to reach a predefined admissible threshold, $\Delta F_{nadir}$, can be estimated according to \eqref{equ2} (see \cite{sigrist2014sizing}).  

\begin{equation} \label{equ2}
    {2H_{sys,gg} \cdot \hat K_{sys,gg} \cdot \Delta F_{nadir}^2 \cdot S_{base}^2 =  P_{crit,gg,t}^2\;\;\;\text{\scriptsize $\;gg\in\mathcal{G}$}} 
\end{equation}

where

\begin{equation}
    H_{sys,gg} = \sum\limits_{g\in\mathcal{G}, g\ne gg}{H_g \cdot x_{g,t}} 
\end{equation}
\begin{equation}
    \hat K_{sys,gg} = \sum\limits_{g\in\mathcal{G}, g \ne gg}{\frac{K_g}{T_g} \cdot x_{g,t}}
\end{equation}

If the outage of a generation unit, $L_{gg} \cdot P_{gg,t}$, is larger than $P_{crit,gg,t}$, load shedding takes place. Conversely, if the outage is smaller than $P_{crit,gg,t}$, no load shedding is needed. If the $\Delta F_{nadir}$ is sufficiently large, the critical unbalance is always larger than the outage of each generation unit and consequently, no load shedding occurs. Ideally, the amount of shed load, $P_{ufls,gg,t}$, is equal to:

\begin{multline} \label{equ5}
    P_{ufls,gg,t} = \begin{cases}
    {L_{gg} \cdot P_{gg,t} -  P_{crit,gg,t}} \\ 
    \text{\;\;\;\;\;\;\;$if$ } {L_{gg} \cdot P_{gg,t} > P_{crit,gg,t}}\;\text{\scriptsize $gg\in\mathcal{G}$} \\
    0\text{ $else$}
    \end{cases}
\end{multline}

This ideal value of $P_{ufls,gg,t}$ is a good (rough) approximation for advanced (conventional) \ac{ufls} schemes as proposed in the literature \cite{sigrist2016island}. 

\Cref{equ2} holds true as long as generation output limits are not hit. This assumption is guaranteed by imposing the following constraint on the absolute generation output, enabling each generation unit $g$ to provide the required power during the transient.

\begin{equation} \label{equ7}
    {P_{g,t} + \frac{{\hat K}_g \cdot x_{g,t}}{\hat K_{sys,gg}} \cdot P_{crit,gg,t} \le \overline{\mathcal{P}}_{g,t} \cdot x_{g,t}\;\;\;\text{\scriptsize $\;g\in\mathcal{G},\;gg \ne g$}} 
\end{equation}

Equations \eqref{equ2} to \eqref{equ7}  need to be linearized and added to the standard \ac{uc} formulation in section \ref{sec:stuc} for the outage of each generation unit $gg$. Well-known techniques exist to linearize products of binary variables or continuous variables and products of binary and continuous variables (e.g., \cite{ferrandon2022inclusion}). Finally, equations \eqref{equ1} and \eqref{equ1k} need to be modified to account for the potential \ac{ufls} as follows:

\begin{equation}
    {\min\limits_{x,P} suc(x_{g,t})+gc(P_{g,t})+lsc(x_{g,t},P_{g,t})}
\end{equation}



\begin{multline}
        \sum\limits_{g\in\mathcal{G}, g\neq gg}\Big(\overline{\mathcal{P}}_{g,t} \cdot x_{g,t}-P_{g,t}\Big)\geq \\
        L_{gg} \cdot P_{gg,t} - P_{ufls,gg,t}\;\;\;\text{\scriptsize $gg\in\mathcal{G}$}
\end{multline}

Note that the cost of \ac{ufls}, $lsc(.)$, is computed by assuming a parameter, $C_{ufls,gg}$ (i.e., the cost of $P_{ufls,gg,t}$ is equal to $C_{ufls,gg} \cdot P_{ufls,gg,t}$). Since a disturbance leading to \ac{ufls} only occurs with some probability, $\rho_{gg}$, $C_{ufls,gg}$ should weight the \ac{vll} by this probability (i.e., $C_{ufls,gg} = \ac{vll} \cdot \rho_{gg}$) because the \ac{cfcuc} contemplates the disturbances (loss of generating units) as certain. Indeed, larger values of $C_{ufls,gg}$ favor a preventive re-dispatch of generating units over \ac{ufls}, and the proposed \ac{cfcuc} becomes a \ac{pfcuc}.

\section{Application to a Spanish island power system}

\label{sec:results}

This section studies the impact of the proposed \ac{cfcuc} formulation on the operation planning of a Spanish island power system. This power system consists of 11 conventional generation units and it has a peak demand of around 35 MW. About $6\%$ of the installed generation capacity currently belongs to wind power generation; \acs{res} covers about $10\%$ of the yearly demand. 
\Cref{tab:techparam} shows the technical parameters of the 11 generating units.


\begin{table}[!ht]
    \caption{Technical Parameters of the Generating Units}
    \centering
    \begin{tabular}{p{0.05\linewidth}|p{0.1\linewidth}|p{0.1\linewidth}|p{0.1\linewidth}|p{0.06\linewidth}|p{0.06\linewidth}|p{0.06\linewidth}}
    \hline
        \textbf{Unit} & \textbf{$\overline{\mathcal{P}}$} (MW) & \textbf{$\underline{\mathcal{P}}$} (MW)  & \textbf{$M_{base}$} (MVA) & \textbf{$H$} (s) & \textbf{$K$} (pu) & \textbf{$T$} (s) \\ \hline
        1 & 3.82 & 2.35 &  5.4 & 1.75 & 20 & 8.26 \\ \hline
        2 & 3.82 & 2.35 &  5.4 & 1.75 & 20 & 8.26 \\ \hline
        3 & 3.82 & 2.35 &  5.4 & 1.75 & 20 & 8.26 \\ \hline
        4 & 4.3 & 2.82 &  6.3 & 1.73 & 20 & 8.26 \\ \hline
        5 & 6.7 & 3.3 & 9.4 & 2.16 & 20 & 8.26 \\ \hline
        6 & 6.7 & 3.3 & 9.6 & 1.88 & 20 & 8.26 \\ \hline
        7 & 11.2 & 6.63 & 15.75 & 2.1 & 20 & 8.26 \\ \hline
        8 & 11.5 & 6.63 & 14.5 & 2.1 & 20 & 8.26 \\ \hline
        9 & 11.5 & 6.63 & 14.5 & 2.1 & 20 & 8.26 \\ \hline
        10 & 11.5 & 6.63 &  14.5 & 2.1 & 20 & 8.26 \\ \hline
        11 & 21 & 4.85 &  26.82 & 6.5 & 21.25 & 3.28 \\ \hline
    \end{tabular}
    \label{tab:techparam}
\end{table}

\begin{figure*}[!ht]
\centering
\subfloat[]{\includegraphics[scale=0.21]{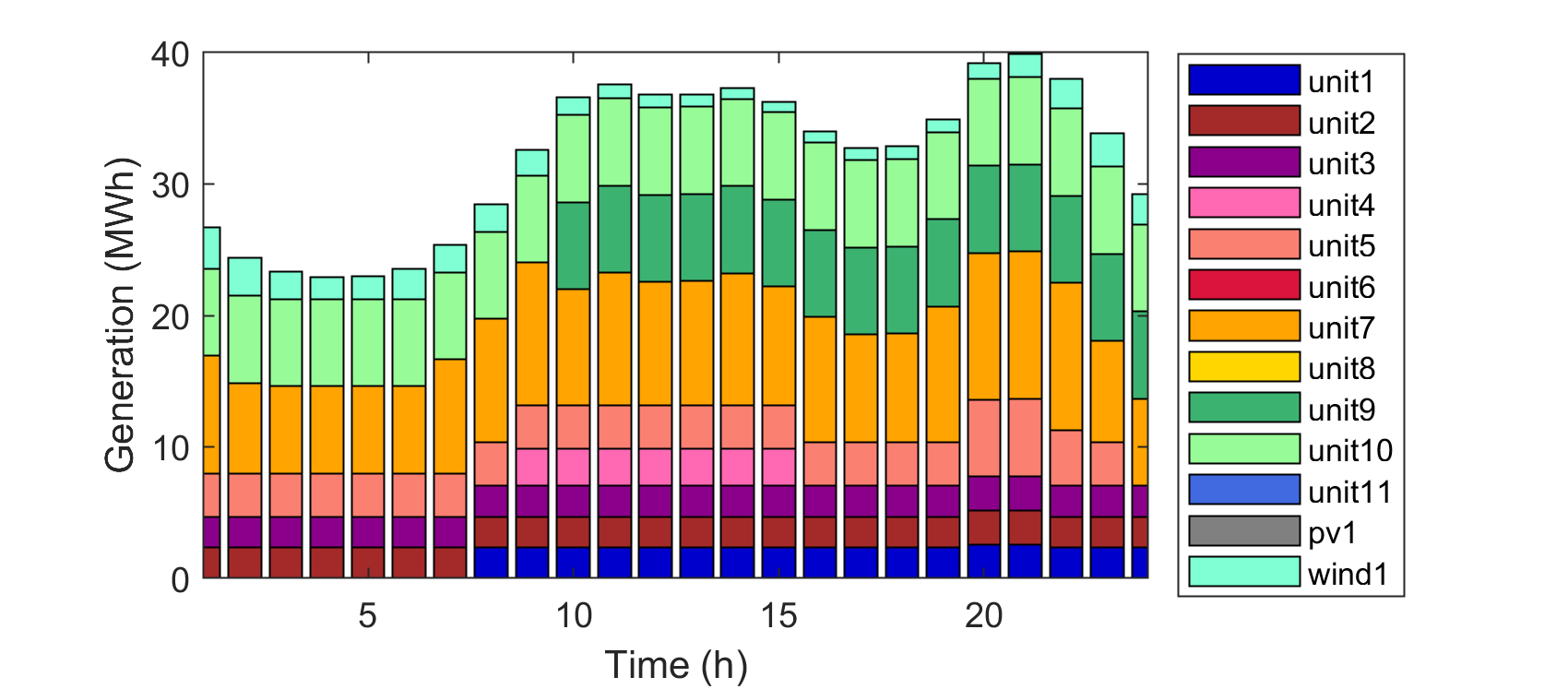}}
\subfloat[]{\includegraphics[scale=0.21]{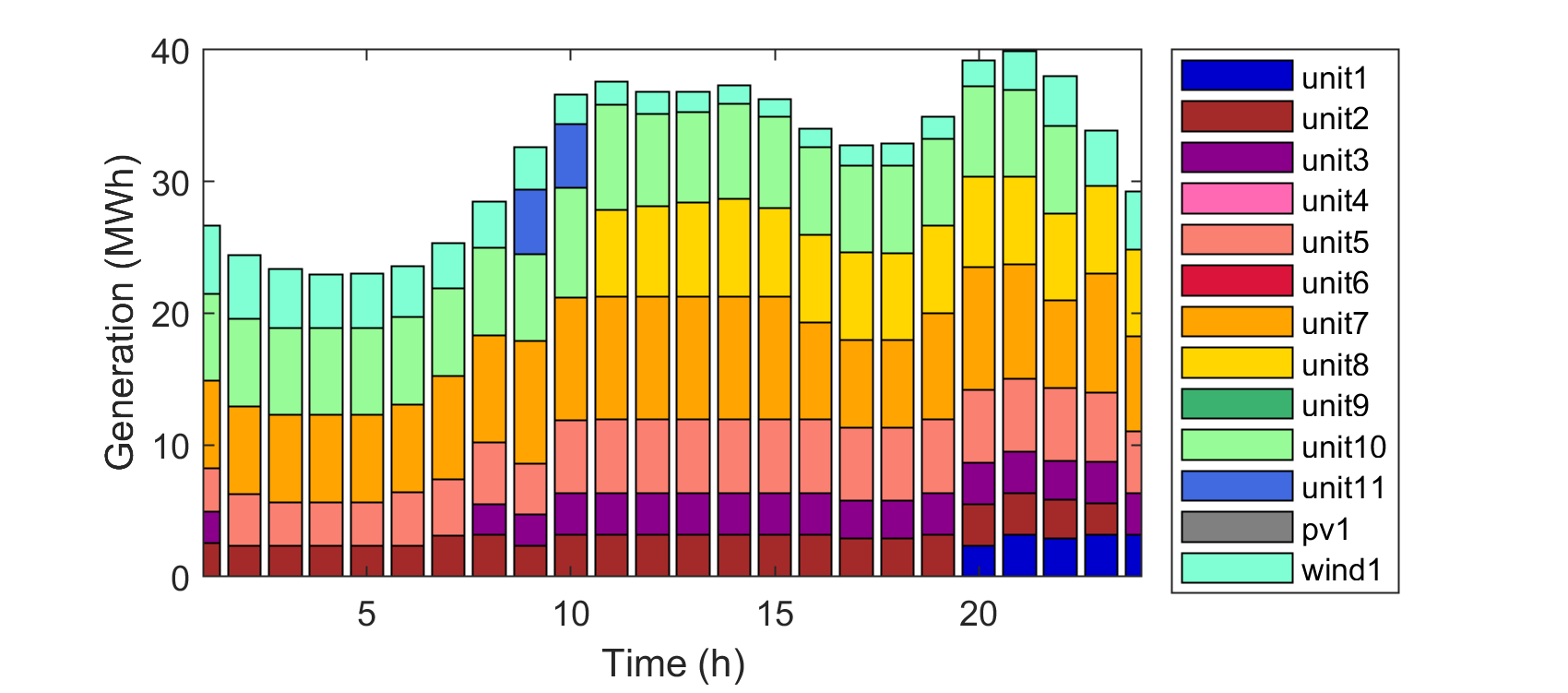}}
\caption{Hourly generation dispatch: (a) Standard \ac{uc} (b) $C_{ufls,gg} = 50$}
\label{fig_dispatchcomp}
\end{figure*}

The considered incidents include the loss of any connected generation unit and the outage of 20\% of \acs{res} generation. $\Delta F_{nadir}$ is set to -2 Hz. The impact of the $C_{ufls,gg}$ will be analyzed.

Fig. \ref{fig_dispatchcomp} shows and compares the generation dispatch resulting from the standard \ac{uc} and the proposed \ac{cfcuc} with $C_{ufls,gg} = 50$. The total system operation costs, the amounts of spilled \acs{res}, and the amount of \ac{ufls} per outage are shown in \Cref{tab:costspillageufls}. The proposed \ac{cfcuc} results in a slightly better dispatch with respect to the standard \ac{uc} in terms of costs. This reduction in system operation costs is due to the consideration of \ac{ufls}, which amounts to about 2 MW per generation outage. Dynamic simulations of all possible generation outages for all generation dispatch scenarios have shown that the standard \ac{uc} leads to an amount of \ac{ufls} of 2.57 MW per generation outage, whereas the proposed \ac{cfcuc} with $C_{ufls,gg} = 50$ leads to an amount of \ac{ufls} of 2.12 MW per generatoin outage, being close to the estimated amount of \ac{ufls} of 2.02 MW per outage.

\begin{table}[!ht]
    \caption{System operation costs, \acs{res} spillage, and \acs{ufls}}
    \centering
    \begin{tabular}{p{0.4\linewidth}|p{0.15\linewidth}|p{0.1\linewidth}|p{0.1\linewidth}}
    \hline
        \textbf{Case} & \textbf{Cost} (\euro) & \textbf{Spillage} (MWh) & \textbf{\ac{ufls}} (MW/out) \\ \hline
        standard \ac{uc} & 140856.1 & 2.8 & - \\ \hline
        \ac{cfcuc}: $C_{ufls,gg} = 50$ \euro & 137023.3 & 0.0 & 2.02\\ \hline
        \ac{cfcuc}: $C_{ufls,gg} = 500$ \euro & 144986.0 & 3.4 & 1.67 \\ \hline
        \ac{cfcuc}: $C_{ufls,gg} = 10^4$ \euro & 152575.7 & 7.9 & 0.0 \\ \hline
    \end{tabular}
    \label{tab:costspillageufls}
\end{table}

Finally, the impact of \ac{ufls} costs is analyzed. Constant $C_{ufls,gg}$ of 50, 500 and of $10^4$ \euro/MW are assumed for this purpose, where the latter cost is sufficiently high to avoid \ac{ufls}. A very large $C_{ufls,gg}$ should actually favor a preventive re-dispatch of generating units instead of \ac{ufls}. \Cref{tab:costspillageufls} shows and compares the impact of non-zero \ac{ufls} costs on system operation costs, \acs{res} spillage, and \ac{ufls}. It can be seen that by increasing $C_{ufls,gg}$, system operation costs increase, whereas \ac{ufls} decreases. For $C_{ufls,gg} =  10^4$ \euro/MW, \acs{res} generation is spilled, but \ac{ufls} is avoided. Corrective \ac{ufls} leads to lower system operation costs and less spillage than preventive re-dispatch of generating units.

\section{Conclusions}
\label{sec:conc}
This letter has presented a \ac{cfcuc} for island power systems. The \ac{cfcuc} implements analytical constraints on \ac{ufls}. According to the considered cost of \ac{ufls} and depending on the inclusion of the \ac{ufls} constraints, the \ac{cfcuc} can be converted into a \ac{pfcuc} or the standard \ac{uc}; therefore, the \ac{cfcuc} is a generalization. The proposed formulation has been successfully applied to a Spanish island power system.

\bibliographystyle{IEEEtran}
\bibliography{References}

\begin{thebibliography}{1}
\providecommand{\url}[1]{#1}
\csname url@samestyle\endcsname
\providecommand{\newblock}{\relax}
\providecommand{\bibinfo}[2]{#2}
\providecommand{\BIBentrySTDinterwordspacing}{\spaceskip=0pt\relax}
\providecommand{\BIBentryALTinterwordstretchfactor}{4}
\providecommand{\BIBentryALTinterwordspacing}{\spaceskip=\fontdimen2\font plus
\BIBentryALTinterwordstretchfactor\fontdimen3\font minus
  \fontdimen4\font\relax}
\providecommand{\BIBforeignlanguage}[2]{{%
\expandafter\ifx\csname l@#1\endcsname\relax
\typeout{** WARNING: IEEEtran.bst: No hyphenation pattern has been}%
\typeout{** loaded for the language `#1'. Using the pattern for}%
\typeout{** the default language instead.}%
\else
\language=\csname l@#1\endcsname
\fi
#2}}
\providecommand{\BIBdecl}{\relax}
\BIBdecl

\bibitem{rajabdorri2022robust}
M.~Rajabdorri, E.~Lobato, and L.~Sigrist, ``Robust frequency constrained uc
  using data driven logistic regression for island power systems,'' \emph{IET
  Generation, Transmission \& Distribution}, vol.~0, no.~0, pp. 1--15, 2022.

\bibitem{ferrandon2022inclusion}
C.~Ferrandon-Cervantes, B.~Kazemtabrizi, and M.~C. Troffaes, ``Inclusion of
  frequency stability constraints in unit commitment using separable
  programming,'' \emph{Electric Power Systems Research}, vol. 203, p. 107669,
  2022.

\bibitem{badesa2019simultaneous}
L.~Badesa, F.~Teng, and G.~Strbac, ``Simultaneous scheduling of multiple
  frequency services in stochastic unit commitment,'' \emph{IEEE Transactions
  on Power Systems}, vol.~34, no.~5, pp. 3858--3868, 2019.

\bibitem{shahidehpour2021two}
M.~Shahidehpour, T.~Ding, Q.~Ming, J.~P. Catalao, and Z.~Zeng, ``Two-stage
  chance-constrained stochastic unit commitment for optimal provision of
  virtual inertia in wind-storage systems,'' \emph{IEEE Transactions on Power
  Systems}, 2021.

\bibitem{sigrist2014sizing}
L.~Sigrist, I.~Egido, E.~L. Migu{\'e}lez, and L.~Rouco, ``Sizing and controller
  setting of ultracapacitors for frequency stability enhancement of small
  isolated power systems,'' \emph{IEEE Transactions on Power Systems}, vol.~30,
  no.~4, pp. 2130--2138, 2014.

\bibitem{sigrist2016island}
L.~Sigrist, E.~Lobato, F.~M. Echavarren, I.~Egido, and L.~Rouco, \emph{Island
  power systems}.\hskip 1em plus 0.5em minus 0.4em\relax CRC Press, 2016.

\end{thebibliography}

\end{document}